\documentstyle[12pt,epsfig]{article}
\def\simgt{\rlap{\lower 3.5 pt \hbox{$\mathchar \sim$}} \raise 1pt \hbox {$>$}}
\def\simlt{\rlap{\lower 3.5 pt \hbox{$\mathchar \sim$}} \raise 1pt \hbox {$<$}}

\catcode`\@=11
\long\def\@makefntext#1{
\protect\noindent \hbox to 3.2pt {\hskip-.9pt
$^{{\ninerm\@thefnmark}}$\hfil}#1\hfill}		

\def\@makefnmark{\hbox to 0pt{$^{\@thefnmark}$\hss}}  

\def\ps@myheadings{\let\@mkboth\@gobbletwo
\def\@oddhead{\hbox{}
\rightmark\hfil\ninerm\thepage}
\def\@oddfoot{}\def\@evenhead{\ninerm\thepage\hfil
\leftmark\hbox{}}\def\@evenfoot{}
\def\sectionmark##1{}\def\subsectionmark##1{}}

\renewcommand{\thefootnote}{\fnsymbol{footnote}}

\newcounter{sectionc}\newcounter{subsectionc}\newcounter{subsubsectionc}
\renewcommand{\section}[1] {\vspace*{0.6cm}\addtocounter{sectionc}{1}
\setcounter{subsectionc}{0}\setcounter{subsubsectionc}{0}\noindent
	{\normalsize\bf\thesectionc. #1}\par\vspace*{0.4cm}}
\renewcommand{\subsection}[1] {\vspace*{0.6cm}\addtocounter{subsectionc}{1}
	\setcounter{subsubsectionc}{0}\noindent
	{\normalsize\it\thesectionc.\thesubsectionc. #1}\par\vspace*{0.4cm}}
\renewcommand{\subsubsection}[1]
{\vspace*{0.6cm}\addtocounter{subsubsectionc}{1}
	\noindent {\normalsize\rm\thesectionc.\thesubsectionc.\thesubsubsectionc.
	#1}\par\vspace*{0.4cm}}

\newcounter{appendixc}
\newcounter{subappendixc}[appendixc]
\newcounter{subsubappendixc}[subappendixc]

\renewcommand{\appendix}[1] {\vspace*{0.6cm}
        \refstepcounter{appendixc}
        \setcounter{figure}{0}
        \setcounter{table}{0}
        \setcounter{equation}{0}
        \renewcommand{\thefigure}{\Alph{appendixc}.\arabic{figure}}
        \renewcommand{\thetable}{\Alph{appendixc}.\arabic{table}}
        \renewcommand{\theappendixc}{\Alph{appendixc}}
        \renewcommand{\theequation}{\Alph{appendixc}.\arabic{equation}}
        \noindent{\bf Appendix \theappendixc #1}\par\vspace*{0.4cm}}

\def\abstracts#1{{

\centering{\begin{minipage}{12.2truecm}\footnotesize\baselineskip=12pt\noindent
	\centerline{\footnotesize ABSTRACT}\vspace*{0.3cm}
	\parindent=0pt #1
	\end{minipage}}\par}}

\renewenvironment{thebibliography}[1]
	{\begin{list}{\arabic{enumi}.}
	{\usecounter{enumi}\setlength{\parsep}{0pt}
\setlength{\leftmargin 1.25cm}{\rightmargin 0pt}
	 \setlength{\itemsep}{0pt} \settowidth
	{\labelwidth}{#1.}\sloppy}}{\end{list}}

\newcounter{itemlistc}
\newcounter{romanlistc}
\newcounter{alphlistc}
\newcounter{arabiclistc}

\newcommand{\fcaption}[1]{
        \refstepcounter{figure}
        \setbox\@tempboxa = \hbox{\footnotesize Fig.~\thefigure. #1}
        \ifdim \wd\@tempboxa > 6in
           {\begin{center}
        \parbox{6in}{\footnotesize\baselineskip=12pt Fig.~\thefigure. #1}
            \end{center}}
        \else
             {\begin{center}
             {\footnotesize Fig.~\thefigure. #1}
              \end{center}}
        \fi}

\newcommand{\tcaption}[1]{
        \refstepcounter{table}
        \setbox\@tempboxa = \hbox{\footnotesize Table~\thetable. #1}
        \ifdim \wd\@tempboxa > 6in
           {\begin{center}
        \parbox{6in}{\footnotesize\baselineskip=12pt Table~\thetable. #1}
            \end{center}}
        \else
             {\begin{center}
             {\footnotesize Table~\thetable. #1}
              \end{center}}
        \fi}

\def\@citex[#1]#2{\if@filesw\immediate\write\@auxout
	{\string\citation{#2}}\fi
\def\@citea{}\@cite{\@for\@citeb:=#2\do
	{\@citea\def\@citea{,}\@ifundefined
	{b@\@citeb}{{\bf ?}\@warning
	{Citation `\@citeb' on page \thepage \space undefined}}
	{\csname b@\@citeb\endcsname}}}{#1}}

\newif\if@cghi
\def\cite{\@cghitrue\@ifnextchar [{\@tempswatrue
	\@citex}{\@tempswafalse\@citex[]}}
\def\citelow{\@cghifalse\@ifnextchar [{\@tempswatrue
	\@citex}{\@tempswafalse\@citex[]}}
\def\@cite#1#2{{$\null^{#1}$\if@tempswa\typeout
	{IJCGA warning: optional citation argument
	ignored: `#2'} \fi}}

 1
 1
 1

\font\ninerm=cmr9

\begin{document}

\begin{flushright}
TTP 95--16 \\
hep-ph/9504345 \\
April 1995 \\
\end{flushright}
\vspace{0.5cm}

\centerline{\normalsize\bf TESTING CP PROPERTIES OF HIGGS BOSONS
\footnote{Invited Talk given at the Ringberg Workshop ``Perspectives for
Electroweak Interactions in $e^+e^-$ Collisions,'' Munich, Feb.\ 5--8,1995.}}
\baselineskip=22pt

\centerline{\footnotesize M.\ L.\ STONG}
\baselineskip=13pt
\centerline{\footnotesize\it Inst.\ Theor.\ Teilchenphysik,
Universit\"at Karlsruhe,
Kaiserstra{\ss}e 12}
\baselineskip=12pt
\centerline{\footnotesize\it D-76128 Karlsruhe, Germany}
\centerline{\footnotesize E-mail: ml@ttpux6.physik.uni-karlsruhe.de}

\vspace*{0.9cm}
\abstracts{Possibilites for measuring the $J^{PC}$ quantum
numbers of the Higgs particle through its interactions with
gauge bosons and with fermions are discussed.
Observables which indicate CP violation in these couplings
are also identified.}

\normalsize\baselineskip=15pt
\setcounter{footnote}{0}
\renewcommand{\thefootnote}{\alph{footnote}}

\section{Introduction}

While the Higgs particle in the Standard Model\cite{higgs} must necessarily
be a scalar state, assigned the external quantum numbers $J^{PC} = 0^{++}$,
the Higgs spectrum in extended models such as supersymmetric theories
may also include pseudoscalar ($J^{PC} = 0^{-+}$) states\cite{gh88}.  This
assignment of the quantum numbers suggests the investigation of
experimental opportunities to measure the parity of the Higgs states.
The experimental observables useful in these measurements are
also useful in studying the question of whether CP violation exists
in the Higgs sector.

Several interesting methods exist to study these problems.  The parity
of the Higgs is reflected in the form of its coupling to fermion and
to gauge-boson pairs, thus providing angular correlations in associated
production of Higgs and one $Z$ boson\cite{barger,tozphi} as well as in
the Higgs decays to gauge-boson\cite{barger,dell89} and
fermion\cite{dell2} pairs.

Another possibility is in the production of neutral Higgs
particles in linearly-polarized photon--photon collisions\cite{gh93}.
The production of scalar particles requires parallel polarization of the two
photons involved, whereas pseudoscalar particles require perpendicular
polarization.

These Higgs production and decay mechanisms are discussed below.
The generic notation $H$ is used for the $0^{++}$ particles and
$A$ for the $0^{-+}$ states.  When mixed states are considered,
the notation $\phi$ is used.

\section{Higgs Production in $e^+ e^- \rightarrow Z \phi$.}

We consider an effective lagrangian which contains the Standard Model
couplings of fermions to the $Z$ and $\gamma$, and study the effects
of the following $\phi Z Z$ couplings:
\begin{eqnarray} \label{leff}
{\cal L}_{ef\! f} = a_Z \, \phi Z^\mu Z_\mu +
	b_Z \, \phi Z^{\mu \nu} \widetilde{Z}_{\mu \nu}
\end{eqnarray}
where $Z_{\mu \nu} = \partial_\mu Z_\nu - \partial_\nu Z_\mu$, and
$\widetilde{Z}_{\mu \nu} \equiv {1 \over 2}
\varepsilon_{\mu \nu \alpha \beta} V^{\alpha \beta}$, with the convention
$\varepsilon_{0 1 2 3} = + 1$.
The term $a_Z$ has the form of the Standard Model $\phi ZZ$ coupling
($a_Z = g_Z m_Z/2$)
and would correspond to a CP-even scalar $\phi$,
while the term $b_Z$ corresponds to a CP-odd pseudoscalar $\phi$.
The presence of both terms indicates that $\phi$ is not a CP eigenstate.
In the most general gauge-invariant dimension-six lagrangian\cite{bw,hisz},
there would be additional CP-even terms.  These have been neglected here
under the assumption that they are suppressed two powers of
some large energy scale relative to $a_Z$ and appear only in interference
terms with the above couplings.

The total cross section for $e^+ e^- \rightarrow Z \phi$ contains no
interference terms ($\sim a_Z b_Z$) and thus no observable CP violation.
Figure \ref{fig:tot_crsect} shows the change in the total cross section
for this process for a small coupling $b_Z$ in addition to the Standard
Model $a_Z$.
A forward-backward asymmetry in the Z scattering angle would be a signal for
CP violation in this reaction.  Such an asymmetry is not only CP-odd
but also CP$\widetilde{\rm T}$-odd\cite{hpzh}, and hence is
proportional to the $Z$ width in the approximation of neglecting
imaginary parts in the effective couplings.
Such an asymmetry occurs when one includes $\phi Z \gamma$ couplings
in the lagrangian (\ref{leff}).
Transverse polarization of the electron beams does not provide
additional information for this reaction, whereas longitudinal
polarization is useful for studying the $\phi Z \gamma$ CP-even
couplings\cite{tozphi}. \\

\vspace{0.1cm}
\begin{figure}
\refstepcounter{figure}
\label{fig:tot_crsect}
\end{figure}
\begin{figure}
\refstepcounter{figure}
\label{fig:Zdst}
\end{figure}
\noindent
\begin{minipage}{0.49\linewidth}
{\footnotesize Fig.~1.~~Cross section for
$e^+ e^- \rightarrow Z H$, for ($\sqrt{s}$,$m_H$) =
(200,60) and (300,150) GeV.  The horizontal solid lines
give the Standard Model values and the dotdashed curves show the
dependence on $b_Z$. }
\end{minipage} \hfill
\begin{minipage}{0.49\linewidth}
{\footnotesize Fig.~2.~~Differential cross sections
(separately normalized) for $e^+ e^- \rightarrow ZH$, ($ZA$, $ZZ$),
with ($\sqrt{s},M_H$) = (500,100) GeV.  The solid (dotdashed,dashed)
line indicates the $ZH$ ($ZA$,$ZZ$) cross section. }
\end{minipage}\\
\vspace{0.1cm}
\begin{minipage}{0.49\linewidth}
\epsfig{file=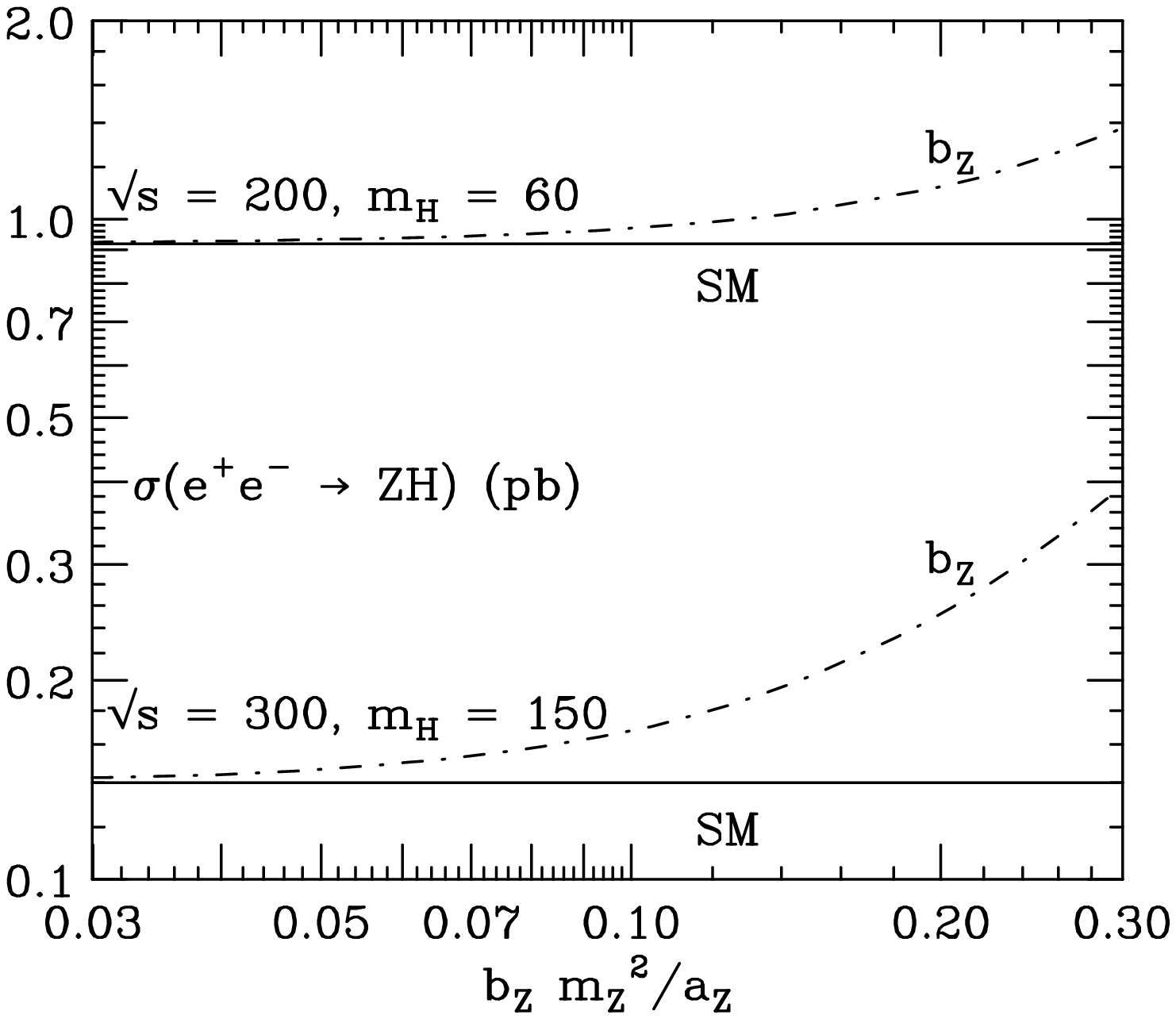,width=1.0\linewidth}
\end{minipage} \hfill
\begin{minipage}{0.49\linewidth}
\epsfig{file=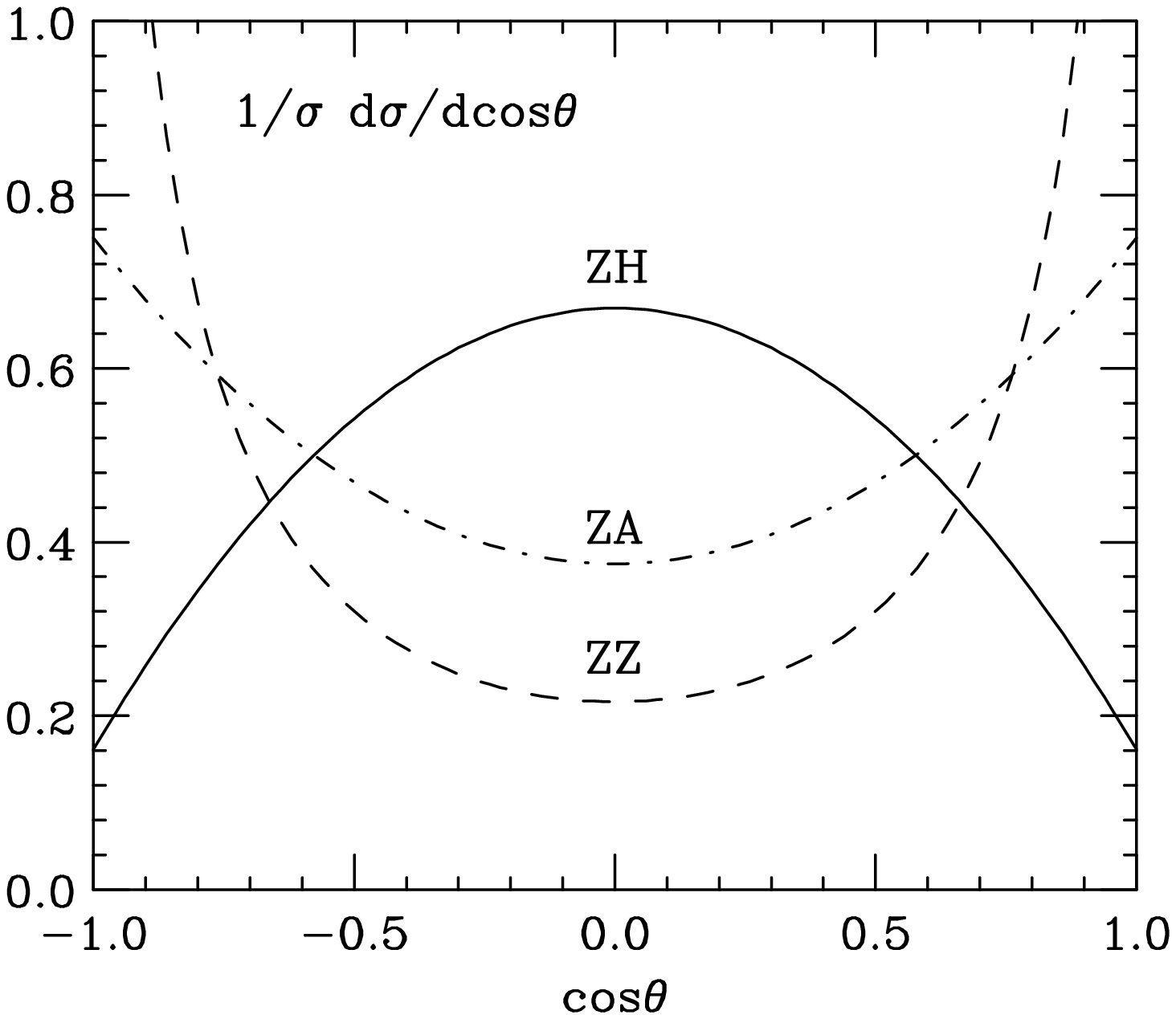,width=1.0\linewidth}
\end{minipage}\\
\vspace{0.1cm}

The two types of terms given here have rather different
characteristics\cite{barger}.
The CP-even term $a_Z$ is an S-wave coupling of the $\phi$ to the $Z$'s.
The $Z$ bosons produced are a mixture of longitudinal and transverse
polarization states.
In the high-energy limit, the mixture becomes purely longitudinal, and
the angular distribution in the $Z$ scattering angle relative to the
electron has the form
$ d\sigma / d \cos \theta \propto \sin^2 \theta $.
The CP-odd coupling $b_Z$ is a P-wave coupling and
the produced $Z$ bosons are purely transversally polarized for any
energy, so that $d \sigma /d \cos \theta \propto  1 - \sin^2 \theta/2$.

Although CP violation may be difficult to observe in this reaction,
the identification of the $\phi$ as a scalar or pseudoscalar should
be possible\cite{barger}.  The background to this identification is the
process $e^+ e^- \rightarrow ZZ$ which is a t-channel process and thus
strongly peaked in large and small $\theta$.  The three angular
distributions are compared in Figure \ref{fig:Zdst}.

More information on the couplings is present in the angular distributions
of the decay products of the $Z$ boson.  The decays of the spin-0
$\phi$ should provide no angular information and are not considered here.
For the $Z$ decay, two more angles are needed to fully describe the
process:  $\hat\theta$ and $\hat\varphi$ are the angles in the $Z$
rest frame between the fermion and the $Z$ boost directions.

The differential cross section for the process $e^+ e^- \rightarrow Z \phi$,
$Z \rightarrow f \bar{f}$ is
\begin{eqnarray} \label{dsig2}
{d \sigma \left( \tau, \tau^\prime \right) \over d \! \cos \! \theta
                \, d \! \cos \! \hat{\theta} \, d \hat{\varphi} } =
        {1 \over 32 \pi s} \bar{\beta} \left({m_Z^2 \over s},
                                        {m_\phi^2 \over s} \right)
{3 {\rm B}(Z \to f \bar{f}) \over 16 \pi}
{ (v_f + \tau^\prime a_f)^2 \over 2(v_f^2 + a_f^2)}
\left| {\cal M}_\tau^{\tau^\prime} \right|^2
\end{eqnarray}
for a given electron helicity $\tau$ and final fermion helicity $\tau^\prime$.
We can expand the squared matrix element above in terms of nine
independent decay angular distributions:
\begin{eqnarray}
\left| {\cal M}_\tau^{\tau^\prime} \right|^2 = &&
	{\cal F}_1 (1 + \cos^2 \hat{\theta}) +
        {\cal F}_2 (1 - 3 \cos^2 \hat{\theta}) +
        {\cal F}_3 \cos \hat{\theta} \nonumber \\ &&
        + {\cal F}_4 \sin \hat{\theta} \cos \hat{\varphi} +
        {\cal F}_5 \sin (2 \hat{\theta}) \cos \hat{\varphi} +
        {\cal F}_6 \sin^2 \hat{\theta} \cos (2 \hat{\varphi}) \nonumber \\ &&
        + {\cal F}_7 \sin \hat{\theta} \sin \hat{\varphi} +
        {\cal F}_8 \sin (2 \hat{\theta}) \sin \hat{\varphi} +
        {\cal F}_9 \sin^2 \hat{\theta} \sin (2 \hat{\varphi}).
\label{fi}
\end{eqnarray}
The distributions are defined such that only the coefficient
${\cal F}_1$ remains after integration over the decay angles
$\hat\theta$ and $\hat\varphi$.

The coefficients ${ \cal F}_i$ may be expressed compactly in terms of
the couplings $a_Z$ and $b_Z$: \\
\begin{tabular}{rlrl}
$ {\cal F}_1 = $ & $ 4 C (1 + \cos^2 \theta ) (s a_Z^2 + 4 p_Z^2 s^2 b_Z^2 )
+ $ & $ 8 C $ & $ \sin^2 \theta e_Z^2 s a_Z^2/M_Z^2$,   \\
$ {\cal F}_2 = $ & $ 8 C\sin^2 \theta e_Z^2 s a_Z^2/M_Z^2 $, &
$ {\cal F}_3 = $ & $ 16 C \tau \cos \theta (s a_Z^2 + 4 p_Z^2 s^2 b_Z^2 )$, \\
$ {\cal F}_4 = $ & $
16 C \tau \tau^\prime \sin \theta E_Z s a_Z^2/M_Z$, &
$ {\cal F}_5 = $ & $ 8 C\sin\theta \cos \theta E_Z s a_Z^2/M_Z$, \\
$ {\cal F}_6 = $ & $ 4 C \sin^2 \theta (s a_Z^2 - 4 p_Z^2 s^2 b_Z^2 )$, &
$ {\cal F}_7 = $ &
$ 32 C \tau \tau^\prime \sin \theta p_Z E_Z s \sqrt{s} a_Z b_Z/M_Z^2$, \\
$ {\cal F}_8 = $ &
$ 16 C \sin\theta \cos\theta p_Z E_Z s\sqrt{s} a_Z b_Z/M_Z^2$, &
$ {\cal F}_9 = $ & $ 16 C \sin^2 \theta p_Z s \sqrt{s} a_Z b_Z$.
\end{tabular} \\
The constant $C = g_Z^2 (v_e + \tau a_e)^2/((s-M_Z^2)^2+M_Z^2\Gamma_Z^2)$
and $E_Z$, $p_Z$ are the energy and momentum of the decaying $Z$ in the
lab.
This result is less than completely general.  In the case that
$\phi Z \gamma$ interactions  are included in the effective lagrangian,
there are more nonzero terms\cite{tozphi}.
In particular, there are CP-violating terms in ${\cal F}_{1,3,4,5}$.
Including the $\phi Z \gamma$ couplings
likewise gives CP-conserving contributions to ${\cal F}_{7,8}$, but
not to ${\cal F}_9$, making this the most interesting of the angular
distribution terms.

We proceed to form asymmetries which isolate the terms above.
First, we integrate out $\theta$, either simply or using a
forward-backward asymmetry:
\begin{eqnarray}
f_i = \int_{-1}^{1} d \cos \theta {\cal F}_i,  \hspace{1cm}
f_i^{\rm FB} = \left( \int_{0}^{1} - \int_{-1}^{0} \right)
d \cos \theta {\cal F}_i,
\end{eqnarray}
then define integrated asymmetries
\begin{eqnarray} \label{aidef}
A_i^{(\rm FB)}(\tau) & = & {1 \over N} \sum_f \sum_{\tau ,\tau ^\prime}
	{3 B(Z \to f \bar{f}) \over 16 \pi }
	{(v_f + \tau^\prime a_f)^2 \over 2 (v_f^2 + a_f^2)}
	f_i^{(\rm FB)}(\tau,\tau^\prime), \\ \label{sigrm}
\sigma_{tot} & = & {1 \over 32 \pi s}
\bar{\beta}({m_Z^2 \over s},{m_\phi^2 \over s}) N.
\end{eqnarray}
These asymmetries are listed in Table \ref{tab:matel}.  In addition,
we indicate which of these asymmetries will be suppressed without
beam polarization or final spin information, and which require
identification of the charge of the final fermion $f$ to be observable.
Figure \ref{fig:zdec_asy} shows the values of the asymmetry $A_9$ for
a small coupling $b_Z$ in addition to the standard coupling $a_Z$. \\

\vspace{0.1cm}
\begin{figure}
\refstepcounter{figure}
\label{fig:zdec_asy}
\end{figure}
\begin{table}
\refstepcounter{table}
\label{tab:matel}
\end{table}
\noindent
\begin{minipage}{0.49\linewidth}
{\footnotesize Fig.~3.~~CP-violating asymmetry $A_9$
(see Eq.~(\ref{aidef})) for ($\sqrt{s}$,$m_H$) = (200,60) GeV (solid) and
(300,150) GeV (dotdashed).
The curves show the dependence on $b_Z$;
when this coupling is zero, the asymmetry is also zero. }
\end{minipage} \hfill
\begin{minipage}{0.49\linewidth}
{\footnotesize Table.~1.~~CP properties of the asymmetries.
CP (non)conservation is indicated with a ($-$)$+$.  The circles
indicate that the charge of the fermion $f$ must be identified, and
the triangles suppression without polarization measurements. }
\end{minipage}\\
\vspace{0.1cm}
\begin{minipage}{0.49\linewidth}
\epsfig{file=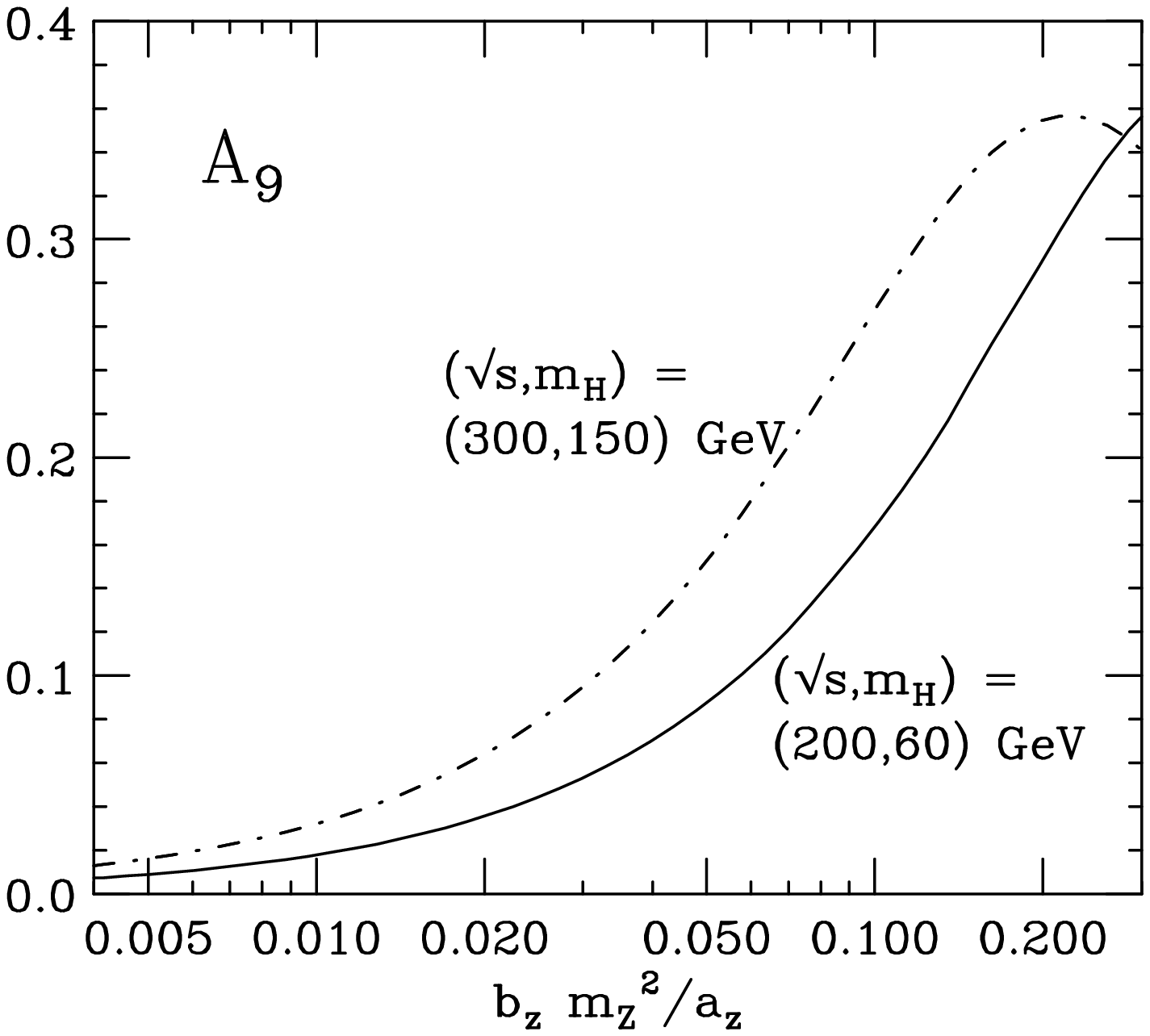,width=1.0\linewidth}
\end{minipage} \hfill
\begin{minipage}{0.49\linewidth}
\begin{center}
\begin{tabular}{c|c|cc|c} \hline
Asym. & CP & beam & $f$  & $f$ \\
           &    & Pol. & Pol. & charge \\ \hline
$\sigma_{tot}$ & $+$ & - & - & - \\ \hline
$A_2$          & $+$ & - & - & - \\ \hline
$A_4$ & $+$ & $\bigtriangleup$ & $\bigtriangleup$ & $\bigcirc$ \\ \hline
$A_5^{\rm FB}\vphantom{A_8^{rm F^B}}$ & $+$ & - & - & - \\ \hline
$A_6$          & $+$ & - & - & - \\ \hline
$A_7$ & $-$ & $\bigtriangleup$ & $\bigtriangleup$ & $\bigcirc$ \\ \hline
$A_8^{\rm FB} \vphantom{A_8^{rm F^B}}$ & $-$ & - & - & - \\ \hline
$A_9$          & $-$ & - & - & - \\ \hline
\end{tabular}
\end{center}
\end{minipage}\\
\vspace{0.1cm}

It is clear that addition of $\phi Z \gamma$ couplings to the effective
lagrangian would produce similar effects in the reaction $e^+ e^- \rightarrow
\phi \gamma$.  Here the Standard Model tree-level coupling does not
exist, so that the small higher-dimension operators would have a
more significant effect, that is, interference effects might be much
larger.  On the other hand, the Standard Model contribution to this
process is tiny\cite{ah1,ah2,ah3} and distributions in the scattering angle
will be more difficult to measure.

\section{Higgs Decays to Vector Bosons}

The decay of Higgs to vector boson pairs provides tests of the Higgs
parity.  The couplings have the same forms in this case as for the
associated production discussed above.  In particular, the
CP-even boson decays to a mixture of longitudinal and transverse
polarization, and the CP-odd Higgs to purely transversally polarized
bosons.
The azimuthal angle $\varphi$ between the vector boson decay planes
may be used to form distributions\cite{dk,nelson,chang}
$d \Gamma(H \rightarrow VV)/ d\varphi \propto 1 + a_1 \cos \varphi
+ a_2 \cos (2\varphi)$,
$d \Gamma(A \rightarrow VV)/ d\varphi \propto 1 - \cos \varphi /4$.
Other useful observables are the energies of the fermions in
the Higgs rest frame\cite{osland,sehgal} and the invariant mass
of the off-shell vector boson\cite{barger} for the decay
$H, A \rightarrow V V^*$.

\section{ $\gamma \gamma$  Production of Higgs Particles}

The colliding photon beam reaction $\gamma \gamma \rightarrow H, A$
has long been recognized (see e.g. Refs.~7,19)
as an important instrument to study the properties of Higgs
particles.
Using linearly polarized photon beams, the parity of
the produced Higgs boson can be measured directly\cite{barger,gg92}.
While the polarization vectors of the two photons must be parallel
to generate scalar particles, they must be perpendicular
for pseudoscalar particles\cite{yang49},
\begin{eqnarray}
{\cal M} (\gamma \gamma \rightarrow H) \sim  \vec \epsilon_1
	\cdot \vec \epsilon_2, \label{eq:amp_phot-to-H}
\hspace{1cm}
{\cal M} (\gamma \gamma \rightarrow A) \sim \vec \epsilon_1
	\times \vec \epsilon_2 \cdot \vec k_\gamma
\label{eq:amp_phot-to-A}
\end{eqnarray}

High-energy colliding beams of linearly polarized photons can be
generated by Compton back-scattering of linearly polarized laser
light on electron/positron bunches of $e^+ e^-$ linear colliders\cite{ginz84}.
The linear polarization transfer from the laser photons to the
high-energy photons is described by the $\xi_3$ component of the
Stokes vector. The length of this vector depends on the final state
photon energy and on the value of the parameter
$x_0 = 4E_e\omega_0/m_{e}^{2}$, where $E_e$ and $\omega_0$ are the
electron and laser energies.
The linear polarization transfer is large for small values of $x_0$
if the photon energy $y = E_\gamma/E_e$ is close to its maximum value.
The maximum value of the Stokes vector $\xi_3(y)$ is reached for
$y = y_{\rm max}$, and approaches unity for small values of $x_0$\cite{kksz}.

Since only part of the laser polarization is transferred to the
high-energy photon beam, it is useful to define the polarization
asymmetry ${\cal A}$ as
\begin{eqnarray}
{\cal A} & = & {N^\parallel - N^\perp \over N^\parallel + N^\perp},
\label{eq:phot_asy_def}
\end{eqnarray}
where $N^\parallel$ and $N^\perp$ denote the number of
$\gamma \gamma$ events with the initial laser polarizations
being parallel and perpendicular, respectively. It follows
from Eq.~(\ref{eq:phot_asy_def}) that
\begin{eqnarray}
{\cal A} (\gamma \gamma \rightarrow H) =
+ {\cal A}, \hspace{1cm}
{\cal A} (\gamma \gamma \rightarrow A) =
- {\cal A}
\label{eq:phot_asy_rels}
\end{eqnarray}

The maximum sensitivity ${\cal A}_{\rm max}$
is reached for small values of $x_0$ and near the upper bound of
$\tau = M_H^2/s_{e^+e^-}$,  i.\ e.\ if the energy is
just sufficient to produce the Higgs particles.
Since the luminosity vanishes at $\tau = \tau_{\rm max}$, the
operating conditions must in practice be set such that a sufficiently
large luminosity is possible.
Typical energies for electron and laser beams are shown in
Table~\ref{tab:x0} for a sample of $x_0$ values corresponding to large
and small asymmetries ${\cal A}_{\rm max}$.
\begin{table}[h,t]
\tcaption{Electron ($E_e$) and laser $\gamma$ energies ($\omega_0$)
for a sample of Higgs masses if the parameter
$x_0 = 4E_e\omega_0/m_e^2$, which determines the maximum
asymmetry ${\cal A}_{\rm max}$ at $\sqrt{\tau_{\rm  max}} =
y_{\rm max}$, is varied from small to large values.} \label{tab:x0}
\begin{center}
\begin{tabular}{|c|cc|c|cc|} \hline
 $x_0$ & ${\cal A}_{\rm max}$ &
 $ \sqrt{\tau_{\rm max}}$ &
 $M_H$~[GeV] & $E_e$~[GeV] & $\omega_0$~[eV] \\
 & & & & (at $\sqrt{\tau_{\rm max}}$) &  \\ \hline\hline
  0.5  & 0.85 & 0.33 & 100 & 150 & 0.22 \\
       &      &      & 200 & 300 & 0.11 \\
       &      &      & 300 & 450 & 0.07 \\ \hline
  1.0  & 0.64 & 0.5  & 100 & 100 & 0.65 \\
       &      &      & 200 & 200 & 0.33 \\
       &      &      & 300 & 300 & 0.22 \\ \hline
  2.0  & 0.36 & 0.67 & 100 & 75  & 1.74 \\
       &      &      & 200 & 150 & 0.87 \\
       &      &      & 300 & 225 & 0.58 \\ \hline
  4.83 & 0.11 & 0.83 & 100 & 60.4& 5.22 \\
       &      &      & 200 & 121 & 2.61 \\
       &      &      & 300 & 181 & 1.74 \\ \hline
\end{tabular}
\end{center}
\end{table}

The measurement of the Higgs parity in $\gamma\gamma$ collisions
will be a unique method in areas of the parameter space where
the Higgs coupling to heavy $W,Z$ bosons are small and the top
quark decay channels are closed so that the Higgs particles
decay preferentially to $b$ and $c$ quarks.
It must therefore be shown that the background events from heavy
quark production can be suppressed sufficiently well.
This is a difficult task\cite{gh93,halzen93} for $b$ quarks.
Three components contribute to the $b \bar{b}$ background events:
direct $\gamma \gamma$ production, the once-resolved photon process
$\gamma\gamma(\to \gamma g) \to b\bar{b}$, and the twice-resolved photon
process $\gamma\gamma(\to gg) \to b\bar{b}$ \cite{gh93,halzen93,drees}.

The cross section for $\gamma \gamma \rightarrow b \bar{b}$
can be easily calculated at the tree level for linearly-polarized
photons. Effects due to higher-order QCD corrections have been shown
to be modest in the unpolarized case\cite{drees} and, hence, can be
safely neglected for asymmetries. The cross sections at energies
sufficiently above the quark threshold are given by
\begin{eqnarray}\label{eq:bb_cr}
{ d \sigma^\parallel \over dy} & = &
{ d \sigma^\perp \over dy}  =
{12 \pi \alpha^2 Q_{b}^{4}  \over s}\:
{ 1 + e^{-4y} \over (1 + e^{-2y})^2 },
\label{eq:b_crsect}
\end{eqnarray}
where $y$ denotes the $b$-quark rapidity.
As evident from Eq.~(\ref{eq:b_crsect}), the background process
$\gamma \gamma \rightarrow b \bar{b}$ does not affect the numerator
of the asymmetry ${\cal A}$, yet it does increase the denominator,
thus diluting the asymmetry in general by a significant amount.
While the signal events are
distributed isotropically in their center-of-mass frame, the
background events are strongly peaked at zero polar angles.
This can be exploited to reject the background events.
In Fig.\ \ref{fig:gg_crsect} we compare the (unpolarized) signal
cross sections in the Standard Model with the
background $b \bar{b}$ channels. \\

\vspace{0.1cm}
\begin{figure}
\refstepcounter{figure}
\label{fig:gg_crsect}
\end{figure}
\begin{figure}
\refstepcounter{figure}
\label{fig:gg_sm}
\end{figure}
\noindent
\begin{minipage}{0.49\linewidth}
{\footnotesize Fig.~4.~~Signal and background cross sections for $b\bar{b}$
final states in the Standard Model.  Here and the subsequent figures,
$m_t = 150$~GeV. }
\end{minipage} \hfill
\begin{minipage}{0.49\linewidth}
{\footnotesize Fig.~5.~~The polarization asymmetry ${\cal A}$ for Standard
Model Higgs production including the background process.}
\end{minipage}\\
\vspace{0.1cm}
\begin{minipage}{0.49\linewidth}
\epsfig{file=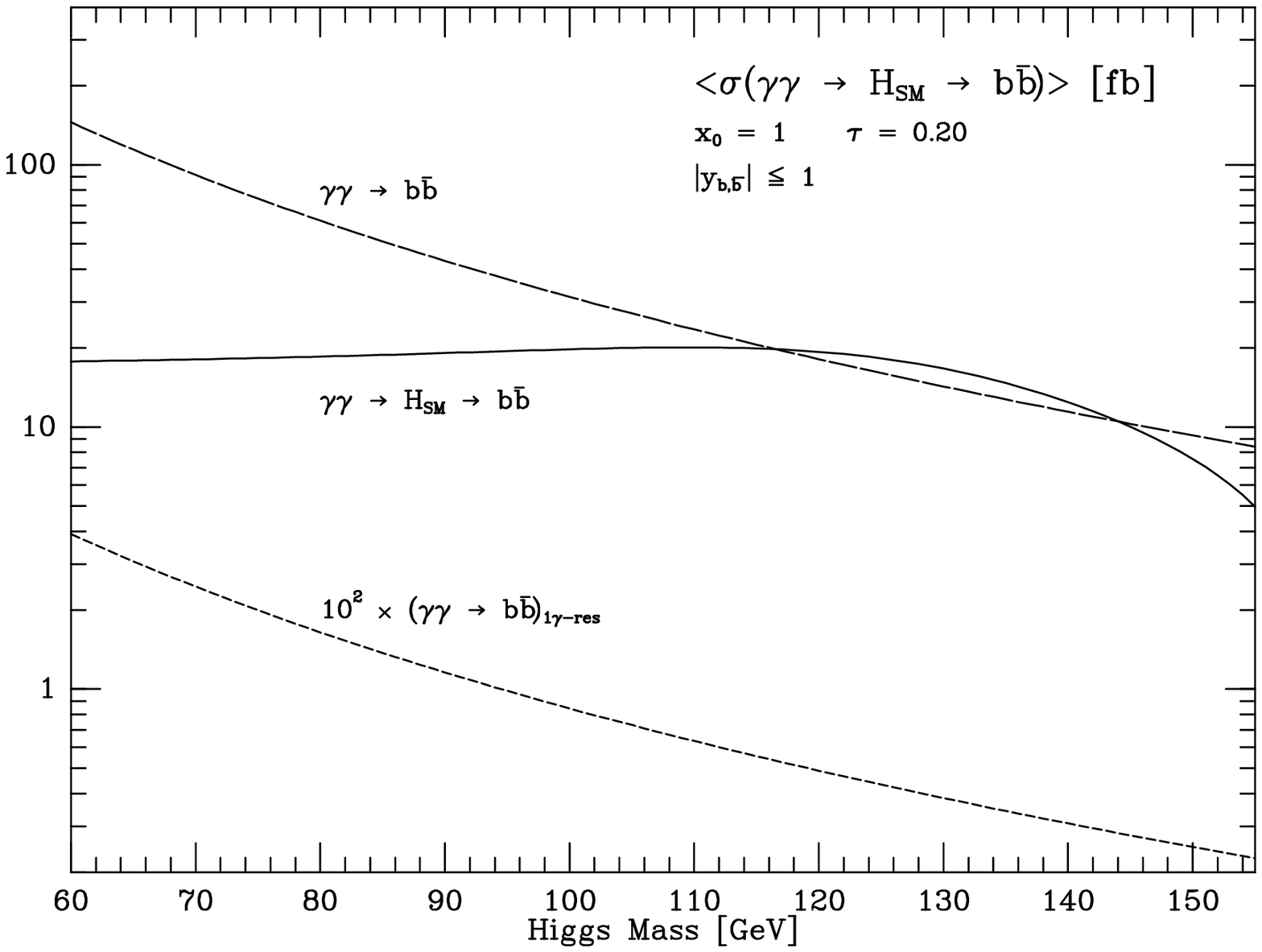,width=1.0\linewidth,bbllx=45pt,bblly=290pt,bburx=565pt,bbury=700pt}
\end{minipage} \hfill
\begin{minipage}{0.49\linewidth}
\epsfig{file=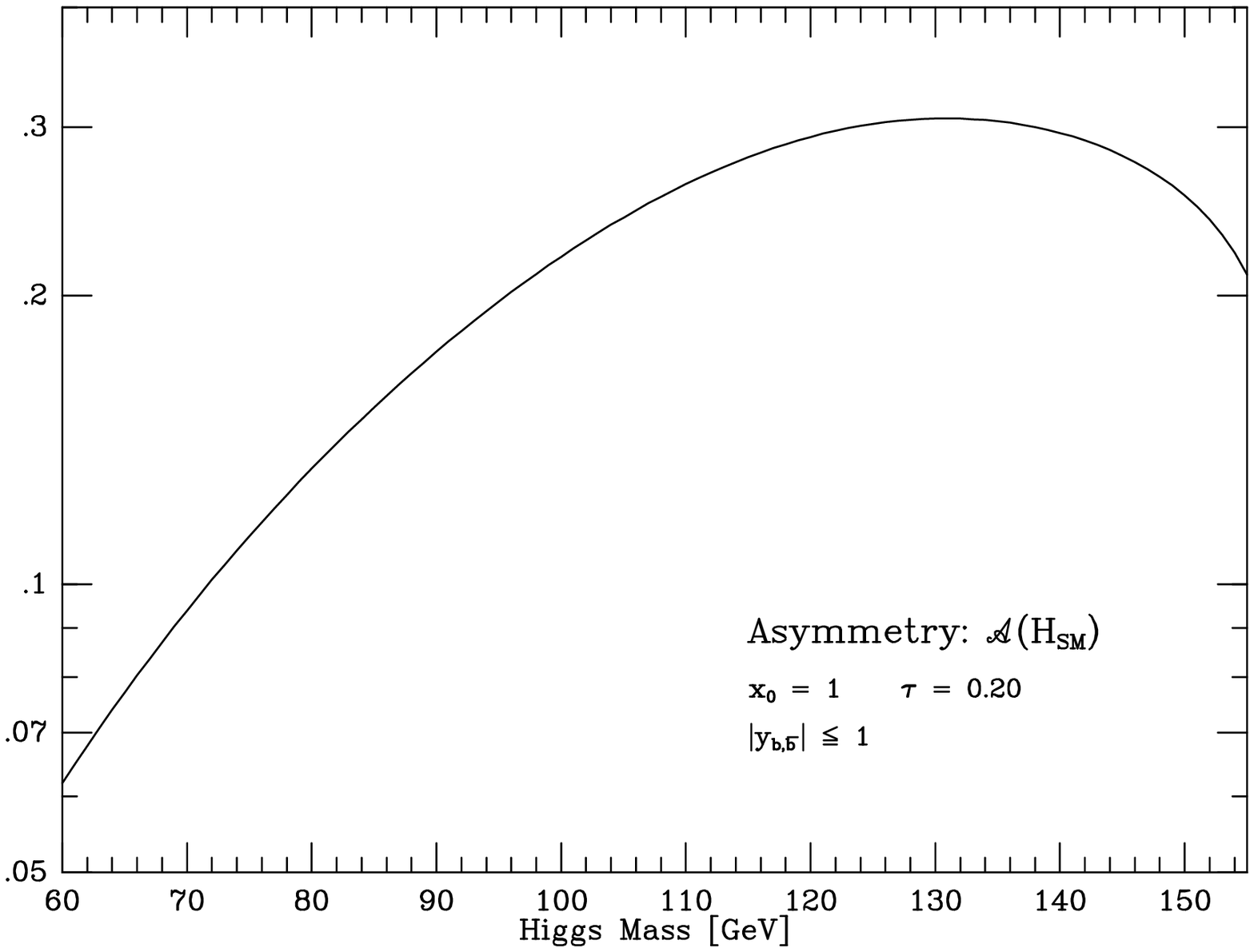,width=1.0\linewidth,bbllx=45pt,bblly=290pt,bburx=565pt,bbury=700pt}
\end{minipage}\\
\vspace{0.1cm}

If one or two photons are resolved into quark-plus-gluon showers,
the subprocesses $\gamma g \rightarrow b \bar{b}$ and $gg \rightarrow
b \bar{b}$ generate $b$-quark final states.  Since gluons are
generated only in the double-splitting process $\gamma \rightarrow q
\rightarrow g$, the gluon spectrum falls off steeply
with gluon momentum.  Therefore, the once- and twice-resolved
processes are strongly suppressed if nearly all the photon energy is
needed to generate the $b \bar{b}$ final state energy.  This is
the situation we encounter for $\tau$ values close to $\tau_{\rm max}(x_0)$,
that is, for large asymmetries ${\cal A}$.
The background from once-resolved processes is thus small in this
kinematical configuration and negligible for the
twice-resolved process.

It is clear from the figures that the
measurement of the Higgs parity, in particular for the heavy particles,
requires high $\gamma \gamma$ luminosities. The background events
reduce the asymmetry ${\cal A}$ by a factor $1/[1 + B/S]$
where $S$ ($B$) denote the number of signal (background) events.
The asymmetries including background events are displayed in
Figs.\ \ref{fig:gg_sm}--\ref{fig:gg_susyc} for the
Standard Model Higgs particle $H_{SM}$ and for the $h^0/A^0$
particles in the minimal supersymmetric model.

The polarization asymmetry of the $\cal{SM}$ Higgs particle
$H_{SM}$  can be measured in $\gamma\gamma$ collisions
throughout the relevant mass range below $\sim 150$~GeV in the
$b\bar{b}$ channel; above this mass value Higgs decays to $Z$ bosons
can be exploited to determine spin and parity.
The light scalar $\cal{MSSM}$ Higgs boson $h^0$ can be probed
in a similarly comprehensive way, except presumably for the low
mass range at large $\tan \beta$.
Finally, the $\gamma \gamma$ polarization measurement of the parity
in the very interesting case of the pseudoscalar $A^0$ Higgs particle
appears feasible throughout most of the parameter range below the
top threshold; $A^0 \rightarrow t \bar{t}$ decays can be exploited
for masses above this threshold.  \\

\vspace{0.1cm}
\begin{figure}
\refstepcounter{figure}
\label{fig:gg_susya}
\end{figure}
\begin{figure}
\refstepcounter{figure}
\label{fig:gg_susyc}
\end{figure}
\noindent
\begin{minipage}{0.49\linewidth}
{\footnotesize Fig.~6.~~The polarization asymmetry ${\cal A}$ for SUSY $h^0$
production including the background process. }
\end{minipage} \hfill
\begin{minipage}{0.49\linewidth}
{\footnotesize Fig.~7.~~The polarization asymmetry ${\cal A}$ for SUSY
$A^0$ production, including background. }
\end{minipage}\\
\vspace{0.1cm}
\begin{minipage}{0.49\linewidth}
\epsfig{file=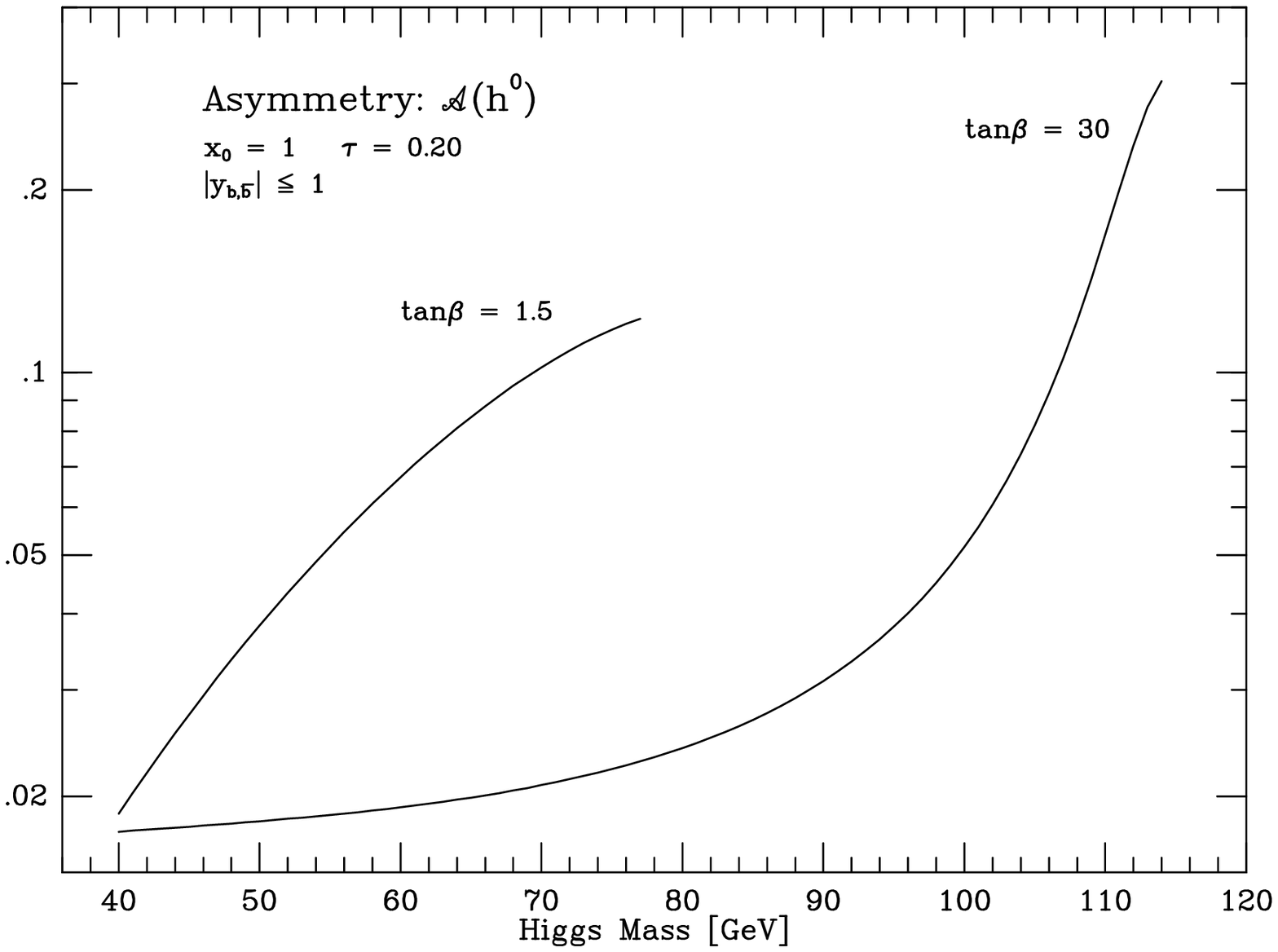,width=1.0\linewidth,bbllx=45pt,bblly=290pt,bburx=565pt,bbury=700pt}
\end{minipage} \hfill
\begin{minipage}{0.49\linewidth}
\epsfig{file=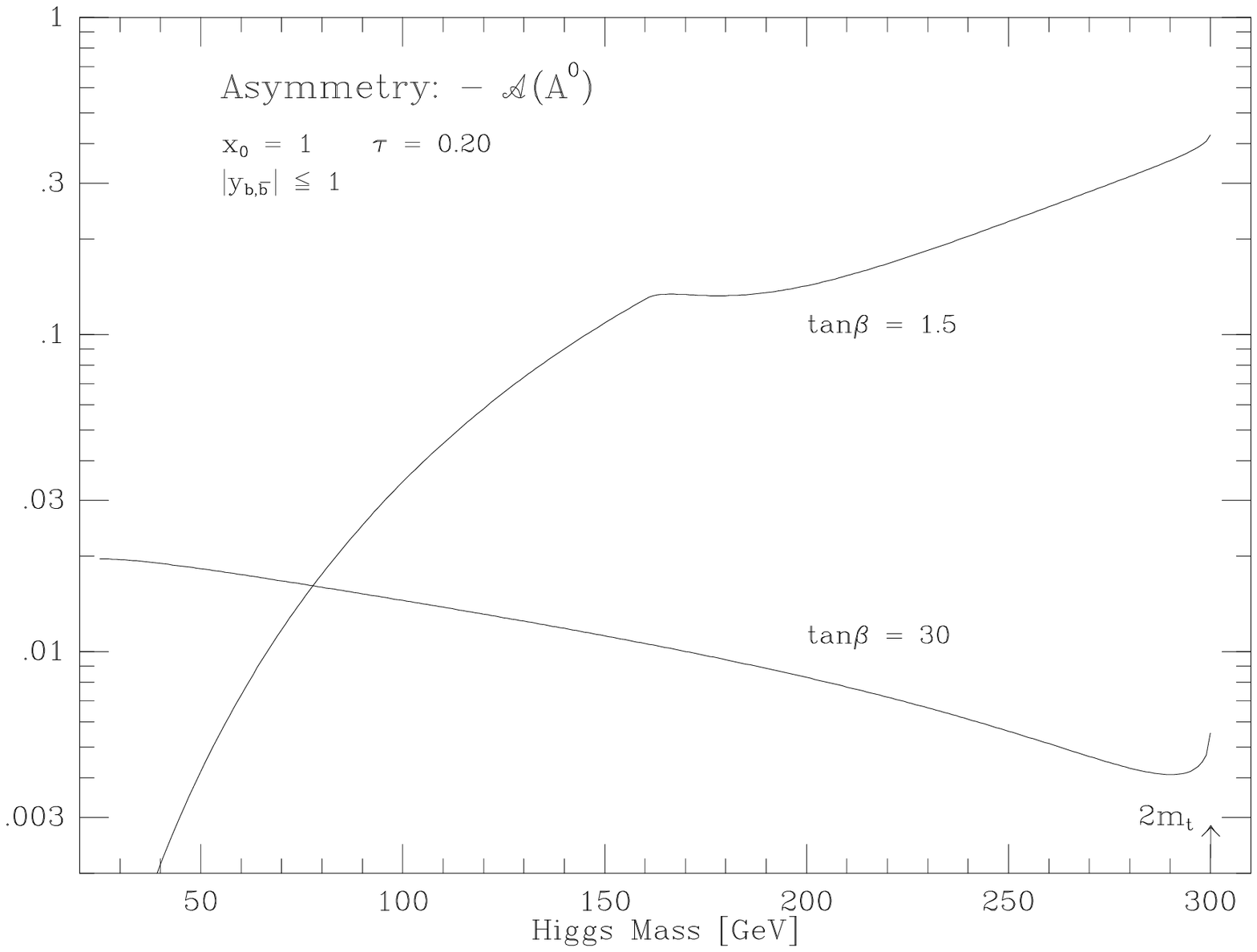,width=1.0\linewidth,bbllx=45pt,bblly=290pt,bburx=565pt,bbury=700pt}
\end{minipage}\\
\vspace{0.1cm}

\section{Neutral Higgs Decays to Fermion Pairs}

The coupling of neutral Higgs particles to fermion pairs also
provides tests of the Higgs parity.  Two conditions on the useful
decay modes exist here.  The first, that it be a mode with relatively
high branching fraction, is satisfied for the $b\bar{b}$ mode and
also for $\tau^+ \tau^-$ and perhaps for $t\bar{t}$.
The $b$ decay channel is in general the most frequent decay mode
in the Standard Model\cite{zerwas1,kniehl} as well as in its minimal
supersymmetric extension\cite{zerwas2}.
Much cleaner channels,
though with branching ratios suppressed by an order of magnitude,
are the $\tau$ and $t$ modes.
The $\tau$ channel is useful in the $\cal{SM}$ for Higgs masses
less than $\sim 130$ GeV and in supersymmetric theories  generally
over a much larger mass range\cite{zerwas2}.  Top quark decays are
of interest in a wide range above the top threshold\cite{ha91}.

The second requirement, that the spin of the fermion be experimentally
observable, is satisfied at present only for the $\tau$ and $t$
decay modes.  Due to the depolarization effects in the
fragmentation process, it is very difficult to extract information
on the $b$ polarization state\cite{mele}.
For large top masses, the top quarks decay
before fragmentation destroys the $t$-spin information\cite{bigi81}.

Denoting the spin vectors of the fermion $f$ and the antifermion
$\bar{f}$ in their respective rest frames by $s$ and $\bar{s}$,
with the $\hat{z}$-axis oriented in the $f$ flight direction, the
spin dependence of the decay probability is given by\cite{barger}
\begin{eqnarray}
\Gamma(H,A \rightarrow f \bar{f}) & \propto & 1 - s_z \bar{s}_z
		\pm s_\perp \bar{s}_\perp  \label{eq:ff_spins}
\end{eqnarray}
This spin dependence translates directly into correlations among
the fermion decay products.

Although the decay mode $\tau^\pm \rightarrow \pi^\pm
\nu_\tau (\overline{\nu}_\tau)$ is rare, it serves as
a simple example.  Defining the polar angles between the $\pi^\pm$ and the
$\tau^-$ direction in the $\tau^\pm$ rest frames by
$\hat\theta_\pm$ and the relative azimuthal angle $\hat\varphi$
between the decay planes, the angular correlation may be
written
\begin{eqnarray}
{1 \over \Gamma} { d \Gamma (H, A \rightarrow \pi^+ \overline{\nu} \pi^- \nu)
\over d \cos \hat{\theta}_+ d \cos \hat{\theta}_- d \hat\varphi} & = &
{1 \over 8 \pi} \left[ 1 + \cos \hat{\theta}_- \cos \hat{\theta}_+
\mp \sin \hat\theta_ + \sin \hat\theta_ - \cos \hat\varphi \right].
\label{eq:pi_dist}
\end{eqnarray}
The full sensitivity to the Higgs parity, reflected in the equal coefficients
of the constant and spin-dependent terms in Eq.~(\ref{eq:ff_spins}),
is retained in this case.
This is a consequence of the spin-0 nature of the pion.

A useful observable sensitive to the parity of the decaying
Higgs particle is the angle $\delta$ between the two charged
pions in the Higgs rest frame\cite{tau_semilept}.
Although the resulting distributions are very similar for most values
of $\delta$, they behave differently in the limit
$\delta \rightarrow \pi$. The scalar distribution approaches
its maximum at $\delta = \pi$, while the pseudoscalar
distribution peaks at a small but nonzero value of $\pi - \delta$.  In
the limit of vanishing pion mass, the distribution approaches
zero as the pions are emitted back-to-back.
Since the pion mass is very much smaller than the $\tau$
mass, the distributions for non--zero pion masses have much
the same behavior in the limit of back-to-back pions.

Other $\tau$ decay modes also provide the opportunity to
extract the Higgs parity.
Let us consider the case of both $\tau$'s decaying to
$\rho ( \widehat{=} 2 \pi )$.  As the $\rho$ is a spin-1 particle, the
correlation term in Eq.~(\ref{eq:pi_dist}),
and hence the sensitivity of the process to the Higgs parity,
is reduced by the factor $(m_\tau^2 - 2 Q^2)^2/(m_\tau^2 + 2 Q^2)^2$.
Predictions for the distribution of the acollinearity
are shown in Fig.\ \ref{f_pidist} for fixed $Q^2 = m_\rho^2$.
The suppression factor is even more severe in the three-pion
channel, where $Q^2 \approx m_\tau^2/2$.

These suppression factors can be avoided in the case that an
event-by-event reconstruction of the $\tau$ decays is possible.
In the $\tau \rightarrow \pi \nu$ decay, the direction of the pion
momentum (defined in the $\tau$ rest frame) appears in Eq.~(\ref{eq:pi_dist}),
replacing the spin vector $\vec{s}$ of Eq.~(\ref{eq:ff_spins}).
In the general case, $s (\bar{s}$) must be replaced by the vector
$\pm R^\mp/(m_\tau \omega_\mp)$, where $R$ and $\omega$ are
defined by\cite{tau_semilept}
\begin{eqnarray}
\Pi_\mu & = & 4 \Re J_\mu q \cdot J^* - 2q_\mu J \cdot J^*
\label{eq:pimu_def}\nonumber \\
\Pi_{5 \mu} & = & 2 \epsilon_{\mu \nu \rho \sigma}
	\Im J^\rho J^{* \nu} q^\sigma
\label{eq:pi5_def}\nonumber \\
\omega & = & p_\mu ( \Pi^\mu - \gamma_{\mbox{{\scriptsize AV}}} \Pi^\mu_5 )
\label{eq:omega_def}\nonumber \\
R_\mu & = & (m_\tau^2 g_{\mu \nu} - p_\mu p_\nu)
	( \gamma_{\mbox{{\scriptsize AV}}} \Pi^\nu - \Pi^\nu_5 )
\label{eq:r_def}
\end{eqnarray}
$q$ is the momentum of the neutrino, $p_\pm$ is the momentum
of the $\tau^\pm$ and $J_\pm$ is the hadronic current, and
$\gamma_{\mbox{{\scriptsize AV}}} = 2 g_{\mbox{{\scriptsize A}}}
g_{\mbox{{\scriptsize V}}} / (g_{\mbox{{\scriptsize A}}}^2
+ g_{\mbox{{\scriptsize V}}}^2)$.
If the $\tau$ rest frame is reconstructed, for example with the
help of microvertex detectors, then $\vec R/(m_\tau \omega)$ can
be evaluated in this frame. Since $ R_\mu R^\mu /(m_\tau \omega)^2 = - 1$,
the sensitivity is completely retained.

For the example of the $\tau$ decay to $\rho \rightarrow 2 \pi$,
the simple distribution in the hadronic momentum gave a reduced
sensitivity to the Higgs parity.  In this case, the optimal
direction for the angular reconstruction\cite{gg_95} is that of the
vector $\vec{R} \propto m_\tau (\vec{\pi}^- - \vec{\pi}^0)
(E_{\pi^-} - E_{\pi^0}) + (\vec{\pi^-} + \vec{\pi^0}) m_\rho^2/2$.
The angles $\hat\theta_\pm$ and $\hat\varphi$ are then those
defining the direction of $\vec{R}$ in the $\tau$ rest frame, and
the angular distribution has the form of Eq.~(\ref{eq:pi_dist}). \\

\vspace{0.1cm}
\begin{figure}
\refstepcounter{figure}
\label{f_pidist}
\end{figure}
\begin{figure}
\refstepcounter{figure}
\label{f_leptdist}
\end{figure}
\noindent
\begin{minipage}{0.49\linewidth}
{\footnotesize Fig.~8.~~Distributions of the decay
$H,A \rightarrow \tau^+ \tau^- \rightarrow h^+ \overline{\nu}_{\tau}
h^- \nu_{\tau}$
in the angle between the hadrons $h = \pi$, $\rho$ for $M_{H,A} = 150$ GeV.
The distributions for scalar (pseudoscalar) Higgs particles are drawn
with solid (dashed) lines. }
\end{minipage} \hfill
\begin{minipage}{0.49\linewidth}
{\footnotesize Fig.~9.~~Distributions of the decays $H,A \rightarrow
t \bar{t} \rightarrow (b l^+ \nu) (\bar{b} l^- \overline{\nu})$ in the angle
between the charged leptons.  Scalar
(solid) and pseudoscalar (dashed) distributions are shown for
$m_t = 150~\mbox{GeV}$ and $m_{H,A} = 400, 1000~\mbox{GeV}$. }
\end{minipage}\\
\vspace{0.1cm}
\begin{minipage}{0.49\linewidth}
\epsfig{file=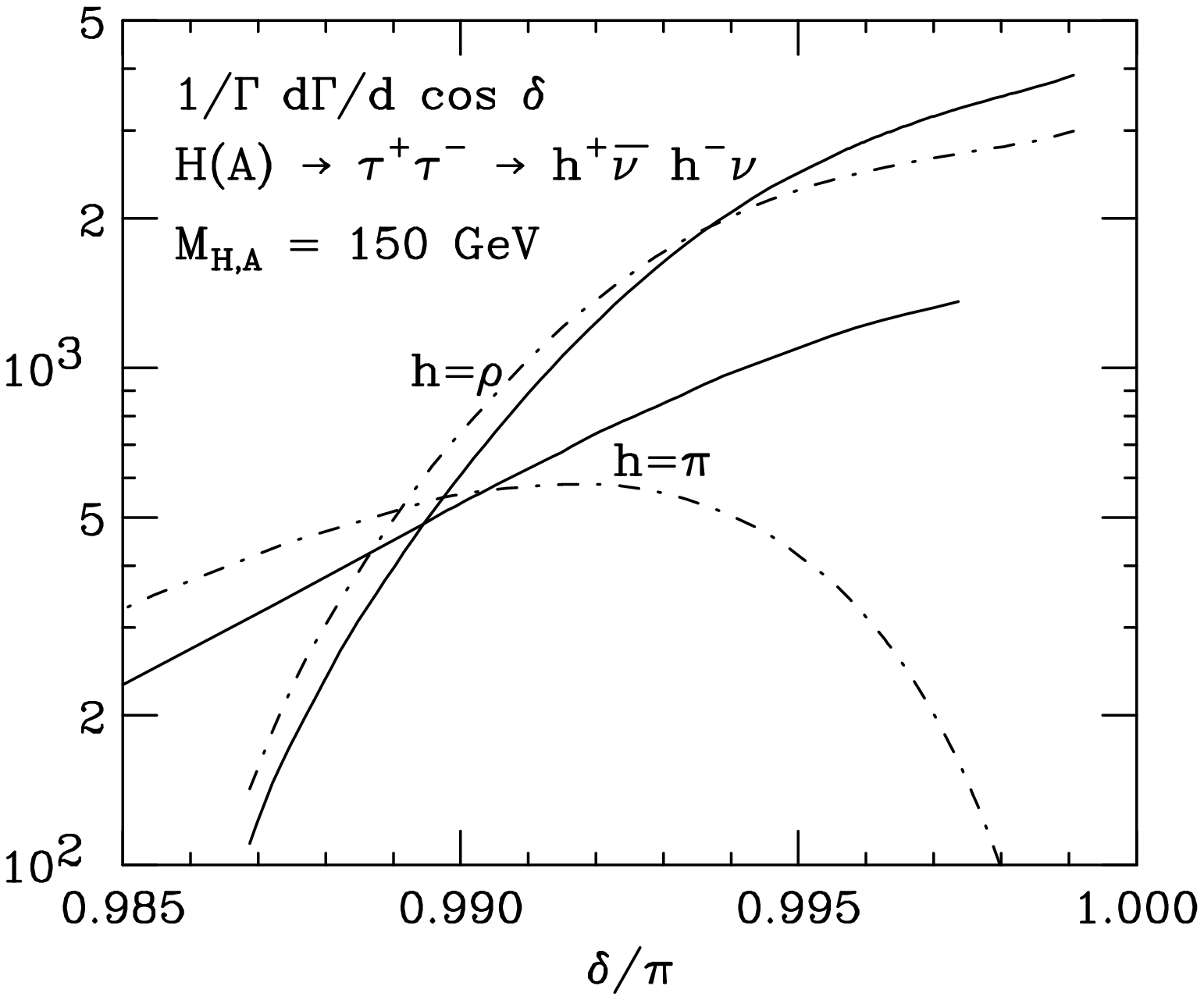,width=1.0\linewidth}
\end{minipage} \hfill
\begin{minipage}{0.49\linewidth}
\epsfig{file=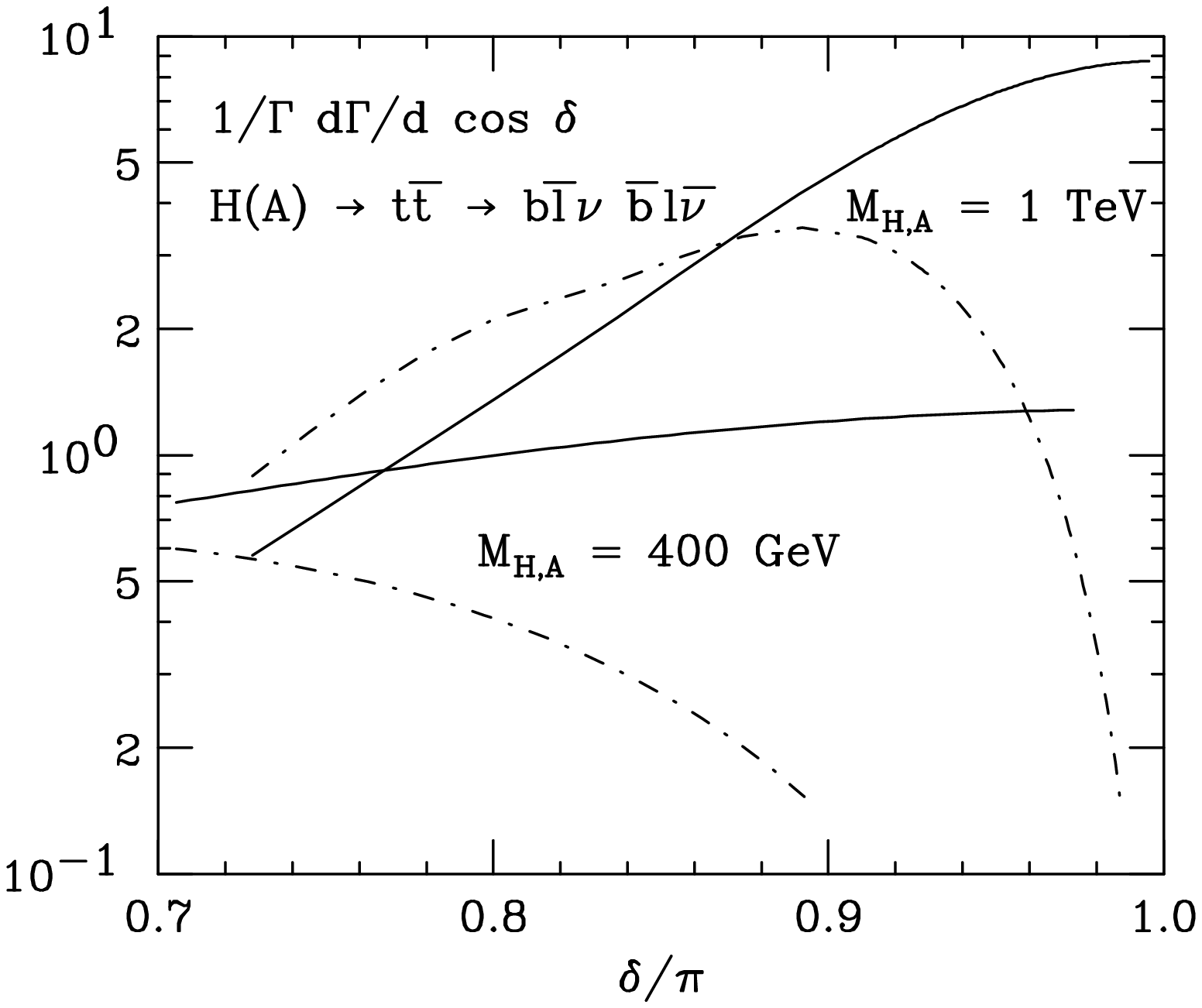,width=1.0\linewidth}
\end{minipage}\\
\vspace{0.1cm}

The decay $H, A \rightarrow t \bar{t} \rightarrow (b W^+)(\bar{b} W^-)$
can be treated in direct analogy to the $\tau$ decay to $\rho$.
In this case, the suppression factor
$\left[ (m_t^2-2 M_W^2)/(m_t^2+ 2 M_W^2) \right]^2 \simeq 0.17$.
A particularly interesting process is provided by subsequent
decays of the $W^\pm$ bosons to leptons.  In this case the
top quark direction can be reconstructed completely. The
distribution obtained after integration over the $b$-quark
directions is exactly the same as in Eq.~(\ref{eq:pi_dist})
with the angles $\hat\theta_\pm$ denoting now the polar angles
between the leptons and the top quarks in the quark rest frames.
Furthermore, the difference between scalar and pseudoscalar
distributions is visible over a much larger angular range,
as the Higgs--to--top boosts are generally small, see
Fig.\ \ref{f_leptdist}.

\section{Conclusions}

The analyses in the preceding sections provide a picture of prospects
to determine experimentally the external quantum numbers $J^{PC}$
of scalar ($H$) and pseudoscalar ($A$) Higgs particles.

The production of Higgs and $Z$ boson
in $e^+ e^-$ reactions is interesting for tests of parity and
of CP violation in the Higgs sector.  The coupling of pseudoscalar Higgs
to vector bosons is however generally not present at tree level,
so that the sensitivity of this process is small for $A^0$.
The situation is analogous for Higgs decay to two vector bosons.

The decay of Higgs to fermion pairs provides angular correlations
sensitive to the Higgs parity for those fermions for which spin
information is experimentally available, the $\tau$ lepton
and, for sufficiently heavy Higgs, the top quark.

The collision of linearly-polarized photon beams is particularly
interesting for tests of the Higgs parity.  The coupling of
both scalar and pseudoscalar Higgs to two photons occurs first
at one-loop level, so that the sensitivity of this process to
both types of states is similar.

\section{Acknowledgements}
I would like to thank K.\ Hagiwara, M.\ Kr\"amer, J.\ K\"uhn, and
P.\ Zerwas for collaborations on the topics discussed in this work.

\end{document}